\documentclass[12pt]{iopart}
\usepackage{iopams}
\usepackage[pdftex]{graphicx}
\usepackage{xcolor,soul}
\begin{document}

\title[A note on Purcell's basic explanation of magnetic
  forces]{A note on Purcell's basic explanation of magnetic forces}

\author{Germano D'Abramo}

\address{Ministero dell'Istruzione, dell'Universit\`a e della Ricerca,
  00041, Albano Laziale, RM, Italy}
\ead{germano.dabramo@gmail.com}
\vspace{10pt}
\begin{indented}
\item[]Dicember 21, 2019
\item[]  
\end{indented}

\begin{abstract}
  In the 1960s, E.M.~Purcell proposed a basic explanation of the magnetic
  force experienced by a test charge moving parallel to a stationary
  current-carrying wire. According to Purcell's derivation, this force results
  from the difference between the relativistic length contraction of the
  distance among the stationary positive charges of the wire and the
  relativistic length contraction of the distance among the negative charges
  moving in the wire when the charges are observed in the rest frame of the
  test charge. The contraction difference generates a charge density imbalance
  that, in the rest frame of the test charge, is experienced as an electrostatic
  force, while, in the lab frame, is perceived as the magnetic force.
  In the present paper, we show that Purcell's approach is problematic since
  it generates inconsistencies and paradoxes. We maintain that Purcell's
  derivation has only an illustrative and expository value and should not be
  taken literally as describing something that really and physically happens
  in the wire. Furthermore, we believe that the difficulties pointed out here
  should be explicitly presented and discussed when introducing Purcell's
  approach in physics courses at the undergraduate and graduate levels.
  
\end{abstract}

%
\vspace{2pc}
\\

\submitto{\EJP}
%
%
%

\section{Introduction}
\label{se1}

The magnetic force felt by a charge moving parallel to a fixed
current-carrying wire is, in fact, an electrostatic force when it is observed in
the rest frame of the charge. E.M.~Purcell provided a basic explanation of this
occurrence~\cite{pur}: in the rest frame of the moving charge, the lattice
positive charges of the wire are seen as moving, and then their density is
modified by Lorentz contraction differently than the density of the
moving electrons, which constitute the current. This charge imbalance is
allegedly at the origin of the magnetic force (i.e., the Lorentz force).

We now explicitly derive Purcell's result referring to the classical
derivations that can be found in the literature~\cite{pur,born,fey,sexl}.
Consider a charge $q$ moving with velocity $v_0$, close by and parallel to a
current-carrying wire at rest in the laboratory system $S$. Let $I$ be the
current flowing in the wire. 
In the $S$-frame, charge $q$ feels only the magnetic (Lorentz) force

\begin{equation}
{\bf F}=q{\bf v_0}\times {\bf B}.
\label{eq1}
\end{equation}

Using Amp\`ere's law, the magnetic field generated by a current $I$ at a
distance $r$ from the current-carrying wire is
$B=\frac{1}{4\pi\epsilon_0 c^2}\frac{2I}{r}$, where $\epsilon_0$ is the vacuum
permittivity, and $c$ is the speed of light in vacuum, and then the
magnitude of the force felts by the charge $q$ is

\begin{equation}
F=\frac{1}{4\pi\epsilon_0 c^2}\frac{2|Iqv_0|}{r}.
\label{eq2}
\end{equation}

The force is repulsive if $q<0$ and $v_0$ is in the same direction as $I$ or
if $q>0$ and $v_0$ is in the opposite direction of $I$.

In the typical atomic description of a metal wire carrying a current, the
electric current comes from the motion of some of the negative electrons
(conduction electrons) while the positive charges of the metal stay fixed.
Let the density of the conduction electrons be $n_{-}$ ($<0$) and their velocity
in $S$ be $v$. The density of the positive charges at rest in $S$ is $n_{+}$,
which is always considered to be equal to the negative of $n_{-}$ since we are
considering an originally neutral wire, and the wire remains uncharged also
when the current passes through it.

Thus, the current $I$ can be written as $n_-vA$, where $A$ is the area of the
cross-section of the wire (notice that the direction of the current $I$ is
opposite to the velocity of the conduction electrons). Then

\begin{equation}
F=\frac{1}{4\pi\epsilon_0 c^2}\frac{2|qn_-Avv_0|}{r}.
\label{eq3}
\end{equation}

Without loss of generality, we take the velocity $v_0$ of the charge $q$ equal
to the velocity $v$ of the conduction electrons, same direction and magnitude.
Thus, with $v_0=v$ eq.~(\ref{eq3}) becomes

\begin{equation}
F=\frac{|q|}{2\pi\epsilon_0}\frac{|n_-|A}{r}\frac{v^2}{c^2}.
\label{eq4}
\end{equation}

If $q>0$, then $F$ is repulsive, while if $q<0$, then $F$ is attractive.

Now let us turn our attention to what happens in the reference frame $S'$,
where the particle is at rest, and the wire is running past with velocity $-v$
(the frame $S'$ moves in the same direction as the conduction electrons and
with the same speed).
The particle is now at rest, so there would be no magnetic force on it. However,
there must be a force, and Purcell's explanation comes into play.

If we take a length $L_0$ of the wire (with cross-section $A_0$), in which there
is a charge density $n_0$ of stationary charges, it will contain a total charge
of $Q=n_0L_0A$. If the same charges are observed in a different frame to be
moving with velocity $v$, they will be all contained in a piece of the
material with the shorter (Lorentz contracted) length

\begin{equation}
  L=L_0\sqrt{1-v^2/c^2},
\label{eq5}
\end{equation}
but with the same area $A_0$ (since, according to special relativity,
dimensions perpendicular to the motion are unchanged). If we call $n$ the
density of charges in the frame in which they are moving, the total charge
$Q$ will be $nLA_0$. This must be also equal to $n_0L_0A_0$ since the charge is
frame invariant so that $nL=n_0L_0$ or, from eq.~(\ref{eq5}),

\begin{equation}
  n=\frac{n_0}{\sqrt{1-v^2/c^2}}.
\label{eq6}
\end{equation}

We can use this result for the positive charge density $n_+$ of our wire. In
the frame $S$, these charges are at rest. In $S'$, however, where the wire moves
with velocity $-v$, the positive charge density becomes

\begin{equation}
  n'_+=\frac{n_+}{\sqrt{1-v^2/c^2}}.
\label{eq7}
\end{equation}

The negative charges, on the other hand, are at rest in $S'$. Their rest
density is $n'_-$, and then

\begin{equation}
  n_-=\frac{n'_-}{\sqrt{1-v^2/c^2}}, \qquad {\textrm{or}}\qquad
  n'_-=n_-\sqrt{1-v^2/c^2}.
\label{eq8}
\end{equation}

We now see that, according to these calculations, there is an electric field
in $S'$ because, in this frame, the net charge density $n'$ is equal to

\begin{equation}
  n'=n'_++n'_-=\frac{n_+}{\sqrt{1-v^2/c^2}}+n_-\sqrt{1-v^2/c^2}
  =n_+\frac{v^2/c^2}{\sqrt{1-v^2/c^2}}>0,
\label{eq9}
\end{equation}
since in $S$, the wire is neutral and $n_-=-n_+$.

The moving wire is thus positively charged and will produce an electric field
$E'$. This field is the same generated by a stationary wire uniformly charged
with linear charge density $\lambda'=n'A$.
The electric field at the distance $r$ from the axis of the wire is

\begin{equation}
  E'=\frac{\lambda'}{2\pi\epsilon_0 r}=\frac{n'A}{2\pi\epsilon_0 r}=
  \frac{n_+A v^2/c^2}{{2\pi\epsilon_0 r\sqrt{1-v^2/c^2}}}.
\label{eq10}
\end{equation}

The electrostatic force acting upon the charge $q$ at rest in $S'$ is then

\begin{equation}
  F'=qE'=\frac{q}{2\pi\epsilon_0}\frac{n_+A}{r}\frac{v^2/c^2}{\sqrt{1-v^2/c^2}}.
\label{eq11}
\end{equation}

Since we have taken the velocity of the frame $S'$ equal to the (drift)
velocity $v$ of the conduction electrons and since $v\ll c$ in ordinary
current-carrying wires, equation~(\ref{eq11}) can be approximated to

\begin{equation}
  F'=\frac{q}{2\pi\epsilon_0}\frac{n_+A}{r}\frac{v^2/c^2}{\sqrt{1-v^2/c^2}}
  \simeq \frac{q}{2\pi\epsilon_0}\frac{|n_-|A}{r}\frac{v^2}{c^2},
\label{eq12}
\end{equation}
which is precisely the magnetic force calculated in $S$ (eq.~(\ref{eq4}))
because of $n_+=|n_-|$.

If we take into account the relativity fact that forces also
Lorentz transform when we go from one system the other, it can
be found that the two ways of looking at the phenomenon do give the
same physical result for any velocity of the particle.

Purcell thus succeeded in showing {\em why} the magnetic (Lorentz)
force experienced by a charge moving parallel to a current-carrying wire is a
purely electrostatic force in the rest frame of the
charge: that force can be ``mechanically'' explained with relativistic
length contraction and the consequent charge density modification in the wire.
No abstract Lorentz transformations of fields are involved in the derivation
of the result.

This explanation appears to be one of the most striking successes of special
relativity: the calculations fit perfectly with the Lorentz force law, and this
seems to be one of the few relativity consequences affecting physical
phenomena in everyday life.

\section{Some inconsistencies deriving from Purcell's explanation}
\label{se2}

There are a few aspects of Purcell's explanation that appear to be problematic.
Inside a metal wire at rest in the laboratory and passed through
by a current, only conduction electrons move (drift). To an observer in the
laboratory, their density should appear augmented with respect to the density 
of the stationary lattice positive charges (which is the same density of 
the electrons when the current is off) by a suitable Lorentz 
factor. Instead, in every quantitative treatment of the problem we have seen,
even when the current is on, in the frame of the wire the charge density of
the moving electrons is considered to be equal to the charge density of the
stationary lattice positive charges ($n_+=|n_-|$) and, to all accounts, the
wire is consistently treated as to be neutral in the laboratory frame.

Unless we are ready to accept that no Lorentz contraction takes place on the
moving electrons with respect to the laboratory frame, we have to explain this
behavior as though, in the laboratory frame, the electrons with a Lorentz
augmented density rearrange themselves in order to restore the zero net charge
that was on the wire before the current was on, more or less according to the
same principle by which negative charges, separated from the positive ones
over a metallic surface, move to cancel out the overall electric potential
difference.

But then the question is: why do electrons not do the same to cancel out the
overall electric potential difference (charge imbalance) also in the reference
frame of the charge moving parallel to the wire? We see in all this a lack of
reciprocity, the same reciprocity demanded by the principle of relativity.
If the relativistic excess of positive charge on the wire, when seen by the
moving charge, is real enough to attract that charge to the wire, then this
same excess of positive charge should be real enough (again, when seen in the
reference frame of the moving charge) to attract more electrons along the wire
from the device that generates the current, in order to reestablish the zero
net charge that was there before the current was on~\cite{prob}.
Why does it not happen? The only possibility is that the wire and the current
generator, when seen from the reference frame of the moving charge, appear
{\em overall} positively charged (non-zero net charge in the whole system).
However, this generates a paradox that shall be discussed more extensively 
in Section~\ref{se3}.

There are few doubts that Purcell's relativistic explanation is quite
impressive. However, can this same relativistic approach explain Amp\`ere's
force law between, for instance, two parallel wires passed through by currents
in opposite directions? Purcell gives an affirmative answer in that
regard. See, for instance, the Example and Fig.~5.23 on pages~263-264 in
\cite{pur}, where Purcell shows how his derivation ``explains the mutual
repulsion of conductors carrying currents in opposite directions''.

We believe that Purcell's approach probably explains why an electron moving in
one of the wires sees a repulsive force exerted by the other wire. If we
consider, for a moment, what an external observer at rest in the
same reference frame of the conduction electrons sees (the observer moves at
the same velocity as the conduction electrons), things become more intricate
in the two-wire case. According to this observer, the wires result overall
oppositely charged. We simply apply the contraction argument to both wires: in
the reference frame of the conduction electrons of a wire, that wire appears
overall positively charged, while the other wire, where  conduction
electrons are seen to be moving faster than the associated positive charges,
appears overall negatively charged. Then, they should attract one another,
exactly the contrary of what Amp\`ere's force law and experience tell us.
The same discrepancy occurs in the case of two parallel wires carrying currents
in the same direction: from Amp\`ere's force law and laboratory experience, we
expect attraction, while Purcell's approach predicts that, in the reference
frame of the moving electrons, both wires become overall positively charged and
then they should repel each other.

If to explain what happens between two current-carrying wires,
we need to resort anyway to Amp\`ere's force law, then the
relativistic explanation of the magnetic force made by Purcell is not general,
and, thus, it cannot be a basic, fundamental explanation of magnetic forces.
Amp\`ere's force is a magnetic force (closely related to the Lorentz force law)
and Purcell's approach cannot explain it with the sole relativistic
modification of the densities of the conduction electrons and the fixed
positive charges, as has been done to explain (successfully) the
Lorentz force law. The exclusive application of Purcell's relativistic
explanation predicts the opposite result, and we always need to resort to
the Amp\`ere force law to match the observations. 

In the following section, we provide a further simple argument against the
interpretation of Purcell's derivation as something that physically
happens in the wire.

\section{Superconducting ring paradox}
\label{se3}

Consider a superconducting ring of radius $R$ in which a direct current $I$
circulates clockwise. In the reference frame of the laboratory, the
ring is uncharged, $Q_{tot}=0$. The ring, when properly cooled, is able to
maintain the current with no applied voltage for years without any
measurable degradation\footnote{The fact that the ring is made of
superconducting material is not essential for the derivation of our main result.
One may also consider a ring made from metal with
extremely low ohmic resistance. In this case, the current will last only for a
few seconds.}. The current generates a magnetic field that
can be very intense. The current $I$ is the result of conduction electrons
moving counterclockwise.

\begin{figure}[t]
\begin{center}
\includegraphics[width=5cm]{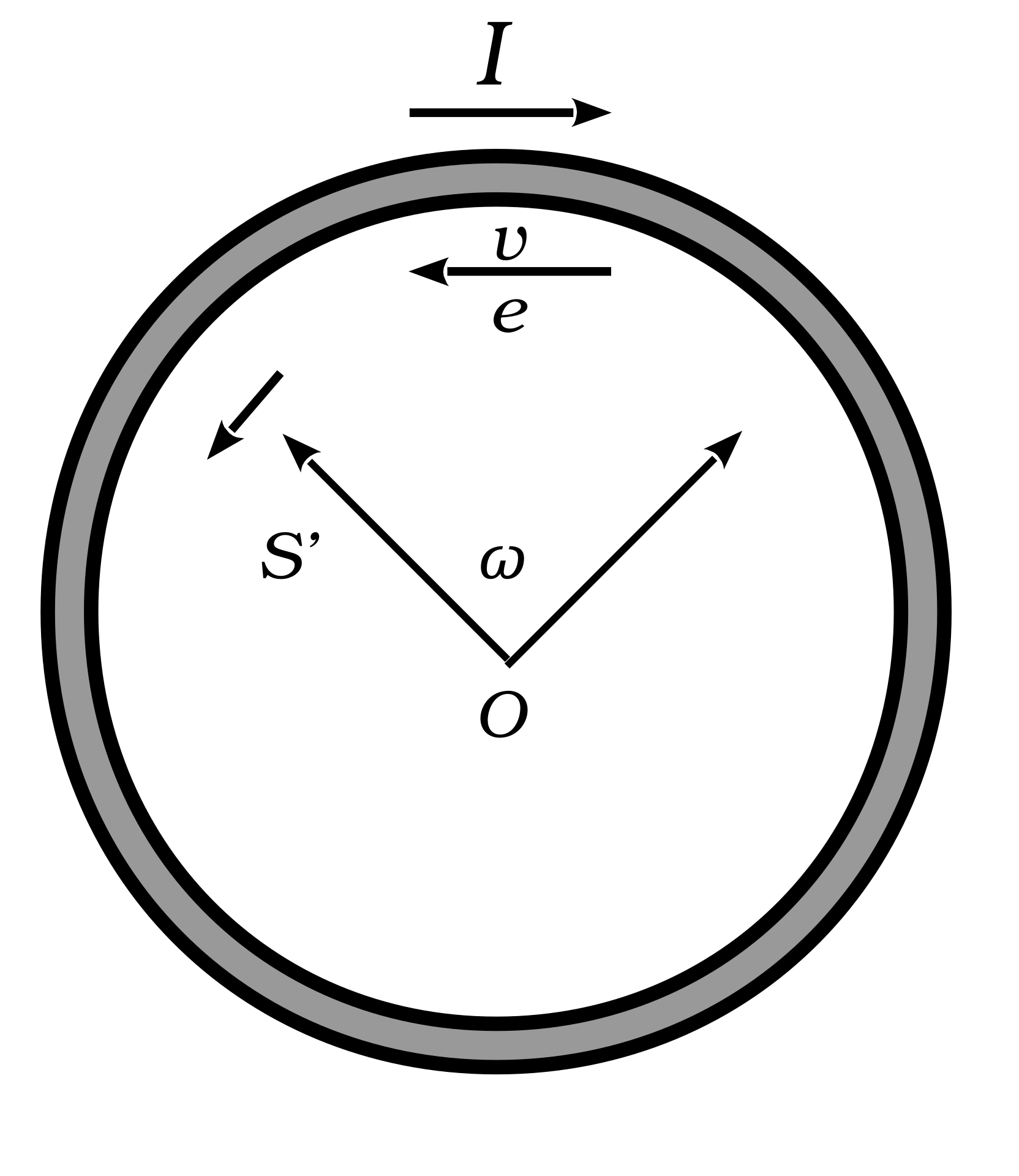}
\end{center}
\caption{Superconducting ring and rotating reference frame $S'$.}
\label{fig1}
\end{figure}

Consider further a reference frame $S'$ with origin $O$ that coincides with
the center of the ring and rotating counterclockwise with an angular velocity
$\omega=v/R$ with $v\ll c$ (Fig.~\ref{fig1}). An observer at rest in $S'$ will
see the conduction electrons as slowed down and the lattice positive charges
of the ring as moving clockwise with velocity $v$. The centripetal acceleration
experienced by $S'$ can be made so small that we can treat this example with
the machinery of special relativity (for instance, for $v<1\,$m/s and
$R\gtrsim 1\,$m $a=\frac{v^2}{R}< 1\,$m/s$^2$). General relativity becomes
special relativity at the limit of a weak field.

If we apply Purcell's approach to the current in the ring, the observer in $S'$
will detect on the whole ring a net positive charge density equal to that
obtained in Section~\ref{se1}, eq.~(\ref{eq9}),

\begin{equation*}
  n'=n_+\frac{v^2/c^2}{\sqrt{1-v^2/c^2}}.
\label{eq13}
\end{equation*}
Here, in some sense, the observer is taking the place of the test particle
described in Section~\ref{se1}.

If we multiply $n'$ by the volume of the ring, we notice that, in $S'$, the ring
is not uncharged, $Q'_{tot}\neq 0$ \footnote{The conclusion does not change if,
in $S'$, we consider a Lorentz contracted ring volume.}.

This result is at odds with the universal principle of charge conservation in
isolated systems. The total charge of the ring must be the same in both
reference frames. What we have just shown also represents a corroboration
of the heuristic observations made in the first part of Section~\ref{se2} about
the electrical neutrality of a rectilinear wire in different reference frames.

To those who still complain that $S'$ is not an inertial frame, and thus
special relativity cannot be applied, consider the following variant. The
reference frame $S'$ is now at rest, and the ring is set in rotation clockwise
with angular velocity $\omega=v/R$. To an observer at rest in $S'$, the
conduction electrons in the ring appear as slowed down, while the lattice
positive charges move clockwise with velocity $v$.
By applying the contraction argument, this time from an inertial reference
frame, the ring appears overall positively charged to the observer in $S'$.
This is again in contradiction with the universal principle of charge
conservation in isolated systems. It is not possible to charge a neutral
current-carrying ring by simply setting it into rotation.

In physics, it is well known that one can generate a potential difference
between the center and the edge of a neutral metallic disk by simply setting
the disk into rotation, but the overall charge of the disk remains always zero
in this case.

\section{Conclusions}
\label{se4}

We have shown that it is not possible to completely (and exclusively) explain
the magnetic force experienced by a charge moving in a static magnetic field
with relativistic length contraction, as has been proposed by E.M.~Purcell in
the case of a field generated by a current-carrying wire.
His explanation results to be at odds with basic physics principles like the
conservation of total charge in isolated systems and even the principle of
relativity.

We maintain that, despite its suggestive agreement with the phenomenological
laws of magnetic forces (the Lorentz force law but not the Amp\`ere force law),
Purcell's basic explanation has only an illustrative and expository value and
cannot be considered as a real physical theory describing what happens in a
current-carrying wire. We believe that the difficulties pointed out here
should be explicitly presented and discussed when introducing Purcell's
approach in physics courses at the undergraduate and graduate levels.

Electric and magnetic fields are interrelated, and this fact is implicit in
Maxwell's equations, but Purcell's relativistic explanation of this
interrelatedness does not work.
The Lorentz transformations of the electric and magnetic fields already tell
that a magnetic field Lorentz transforms into an electric field (and, by
necessity, also into a magnetic field since a purely magnetic field cannot be
transformed into a purely electric field). Here, though, we show that the
electric field, coming from the Lorentz transformation of the purely magnetic
field generated by a current-carrying wire, cannot be simply and
``mechanically'' explained with charge density imbalance in the wire
due to relativistic length contraction.

Our conclusions, although coming from an analysis limited to the very specific
approach of Purcell, agree with the allegedly more general result obtained by
Jefimenko~\cite{je}, according to which neither the magnetic nor the electric
field is a relativistic effect.

\ack
The author is indebted to Dr.~Assunta Tataranni and Dr.~Gianpietro Summa for
key improvements to the manuscript. The author acknowledges anonymous
reviewers for their valuable comments and suggestions.

\end{document}